\PassOptionsToPackage{english}{babel}
\documentclass[reprint,aps,prl,twocolumn,showpacs,amssymb,amsfonts,superscriptaddress,longbibliography,nofootinbib]{revtex4-1}  
\usepackage[english]{babel}
\usepackage{dsfont}
\usepackage{graphicx}  
\usepackage{dcolumn}   
\usepackage{bm}        
\usepackage{bbm}        
\usepackage{amssymb}   
\usepackage{amsmath}
\usepackage{mathtools}
\usepackage[utf8]{inputenc}
\usepackage{siunitx}
\usepackage[amsmath,thmmarks]{ntheorem}
\usepackage{amsbsy}

\usepackage{changes}

\definecolor{myurlcolor}{rgb}{0,0,0.7}
\definecolor{myrefcolor}{rgb}{0.8,0,0}
\usepackage{hyperref}
\hypersetup{colorlinks, linkcolor=myrefcolor,
citecolor=myurlcolor, urlcolor=myurlcolor}

\usepackage{cleveref}
\crefrangeformat{equation}{#3(#1 #4--#5 #2)#6}

\usepackage{comment}
\usepackage{soul} 
\newcommand{\ignore}[1]{}
\usepackage{enumitem}

\newcommand{\eg}{\textit{e.g. }}				
\newcommand{\ie}{\textit{i.e. }}				
\newcommand{\var}[1]{\ensuremath{\Delta^2( #1)}}

\newcommand{\avg}[1]{\left\langle #1 \right\rangle}		
\newcommand{\tr}{\text{Tr}}

\newcommand{\ket}[1]{\ensuremath{|#1\rangle}}
\newcommand{\be}{\begin{equation}}
\newcommand{\ee}{\end{equation}}

\newcommand{\abs}[1]{\ensuremath{\left\vert #1\right\vert}}

\newcommand{\EM}{\ensuremath{\mathcal{E}_{\mathcal{M}}}}

\newcommand{\BSA}{\ensuremath{\mathcal{E}_{\text{BSA}}}}
\newcommand{\GR}{\ensuremath{\mathcal{E}_{\text{GR}}}}
\newcommand{\MBSA}{\ensuremath{\mathcal{M}_{\text{BSA}}}}
\newcommand{\MGR}{\ensuremath{\mathcal{M}_{\text{GR}}}}

\newcommand{\lmax}[1]{\ensuremath{\lambda_{\text{max}}(#1)}}
\newcommand{\lmin}[1]{\ensuremath{\lambda_{\text{min}}(#1)}}

\definecolor{darkgreen}{RGB}{50,190,50}
\definecolor{darkblue}{RGB}{0,0,190}
\definecolor{darkred}{RGB}{238,0,0}
\definecolor{quantum}{RGB}{83,37,127}
\definecolor{quantumlight}{RGB}{169,146,191}

\newcommand{\id}{\mathbbm{1}} 

\begin{document}
\selectlanguage{english}

\title{Entanglement quantification in atomic ensembles}

\author{Matteo Fadel}
\email{matteo.fadel@unibas.ch} \affiliation{Department of Physics, University of Basel, Klingelbergstrasse 82, 4056 Basel, Switzerland} 

\author{Ayaka Usui}
\affiliation{Quantum Systems Unit, Okinawa Institute of Science and Technology Graduate University, Okinawa, Japan}

\author{Marcus Huber}
\affiliation{Vienna Center for Quantum Science and Technology, Atominstitut, TU Wien, 1020 Vienna, Austria}
\affiliation{Institute for Quantum Optics and Quantum Information - IQOQI Vienna, \\ Austrian Academy of Sciences, Boltzmanngasse 3, 1090 Vienna, Austria}

\author{Nicolai Friis}
\affiliation{Institute for Quantum Optics and Quantum Information - IQOQI Vienna, \\ Austrian Academy of Sciences, Boltzmanngasse 3, 1090 Vienna, Austria}

\author{Giuseppe Vitagliano}
\affiliation{Institute for Quantum Optics and Quantum Information - IQOQI Vienna, \\ Austrian Academy of Sciences, Boltzmanngasse 3, 1090 Vienna, Austria}

\date{\today}

\begin{abstract}
Entanglement measures quantify nonclassical correlations present in a quantum system, but can be extremely difficult to calculate, even more so, when information on its state is limited. Here, we consider broad families of entanglement criteria that are based on variances of arbitrary operators and analytically derive the lower bounds these criteria provide for two relevant entanglement measures: the best separable approximation (BSA) and the generalized robustness (GR). This yields a practical method for quantifying entanglement in realistic experimental situations, in particular, when only few measurements of simple observables are available. As a concrete application of this method, we quantify bipartite and multipartite entanglement in spin-squeezed Bose-Einstein condensates of $\sim 500$ atoms, by lower bounding the BSA and the GR only from measurements of first and second moments of the collective spin operator.
\end{abstract}

\maketitle

{\it Introduction.---} Entanglement is a form of quantum correlations that constitutes an essential resource for a number of quantum information tasks~\cite{HorodeckiRMP09}. Formally, it is defined as the impossibility of expressing the state of a composite system as the (convex combination of) product of subsystems' states. While this formal definition captures an essential difference between classical and quantum systems, deciding whether a given experimental quantum system exhibits entanglement is not an easy task. Nevertheless, a method for entanglement detection that is often successful employs entanglement witnesses~\cite{GuheneToth09, FriisNatPhys19}. These are observables represented by Hermitian operators $W$ such that $\tr[W\sigma]\geq 0$ for all separable states $\sigma$. Therefore, observing $\tr[W\rho]< 0$ implies that $\rho$ is entangled.

While entanglement witnesses allow us to certify that a state is entangled, and even to characterize its separability structure, they do not provide any direct information on the ``strength'' of these correlations, for example in terms of robustness to noise. In other words, observing, \eg, $0>\tr[W\rho_1]>\tr[W\rho_2]$, or that $\rho_2$ involves more entangled particles than $\rho_1$, does not necessarily imply that $\rho_2$ is ``more entangled'' than $\rho_1$. To reach this conclusion, a correct quantification of the nonclassical resources present in the states is required.

Here, we consider  entanglement measures to be nonnegative real functions $\mathcal{E}(\rho)$ such that~\cite{VedralPlenio98,PlenioVirmani07}: i) $\mathcal{E}(\sigma)=0$ for all separable states $\sigma$; and ii) $\mathcal{E}(\rho)$ does not increase on average under local operations and classical communication (LOCC)\footnote{We note that this differs from the convention in, e.g.,~\cite{PlenioVirmani07}, where such quantities are referred to as `monotones'.}. 
Many inequivalent measures can be defined, which result in different orderings of the entangled states. For this reason, it is usually favorable to consider measures that are associated to questions of practical relevance, such that they inherit a concrete meaning.

In the bipartite case, typically adopted measures are entropies, Schmidt rank, concurrence or entanglement of formation/distillation \cite{PlenioVirmani07,FriisNatPhys19}. In the multipartite case, however, even more possibilities arise, reflecting the complexity of multipartite LOCC classification and the lack of a unique maximally entangled state \cite{DeVicenteSpeeKraus2013,SauerweinWallachGourKraus2018}. These measures are often operationally related to communication tasks, where entanglement is seen as a resource for transcending the limitations of LOCC for distant parties. For many-body systems, however, other features are typically more relevant.
One measure that is important in this context is the best separable approximation (BSA)~\cite{LewSanp98,KarnasLew01}, which captures to what extent the state $\rho$ of a many-body system can be approximated by a separable quantum state. Formally, the $\BSA$ is defined as the minimum real number $t\in[0,1]$ such that $\rho$ can be decomposed as 
\begin{equation}\label{eq:defBSA}
\rho = (1-t) \,\sigma + t\, \delta \rho \;,
\end{equation}
for some separable density matrix $\sigma$ and some remainder density matrix $\delta \rho$. Another measure relevant in the context of many-body experiments is the generalized robustness (GR)~\cite{SteinerPRA03}, which quantifies the minimal amount of noise (represented by a general state) that needs to be mixed with $\rho$ in order to make it separable. Formally, $\GR$ is defined as the minimum real number $s\in[0,\infty)$ such that
\begin{equation}\label{eq:defGR}
\frac 1 {1+s} \,\rho + \frac s {1+s} \,\rho^\prime \quad \text{is separable} \;,
\end{equation}
where $\rho^\prime$ is any (not necessarily separable) density matrix. Besides quantifying the robustness to noise of an entangled state, the GR is of interest as it also has direct connections to the maximum fidelity of teleportation and other entanglement measures (\eg  entropic monotones, geometric distances)~\cite{brandao05,BandaoVianna06,CavalcantiPRA06}.

Evaluating an entanglement measure $\mathcal{E}(\rho)$ is \textemdash\ in such cases where this is even possible at all \textemdash\ a demanding task, as it requires the full knowledge of $\rho$. To circumvent this requirement, which in most experimental situations is just impossible to fulfill, methods have been developed to at least lower bound interesting measures from limited information on $\rho$~\cite{Audenaert06,EisertBrandaoAudenaert07,GuhneReimpellWernerPRL07,GuhneReimpellWerner08,GittsovichPRA10,MartyPRL16,MartinPRL17,ErkerQuantum17,BavarescoNatPhys18,SchneelochPRA18, FadelFidPRA20,Herrera-ValenciaSrivastavPivoluskaHuberFriisMcCutcheonMalik2020, bergh21}. Although some of these have been applied with extraordinary success in optical experiments (see Ref.~\cite{FriisNatPhys19} for a recent review), finding an approach suitable to platforms where the number of accessible
observables is particularly limited (\eg~collective properties) remains extremely challenging.

In the case of atomic ensembles, entanglement between particles is routinely detected and characterized through criteria based on low moments of collective spin observables~\cite{Lucke2014Detecting,VitaglianoPlanar,PezzeRMP2016,BaccariDepth,FadelGessnerPRA20}. Moreover, bipartite entanglement has also been demonstrated between spatially separated atomic ensembles~\cite{JulsgaardNATURE2001,Fadel18,KunkelEtAl2018,Lange18}. In these systems, however, entanglement quantification has so far been limited to theoretical investigations~\cite{StocktonEntQuantPRA03,KrammerPRL09,CramerPlenioWunderlich11,jing2019}, or to experiments with spinless bosons in optical lattices that assumed superselection rule or purity of the state~\cite{CramerEtAl2013,Islam2015}, \ie~ highly idealised situations.

In this work we leverage the strength of simple many-body entanglement witnesses to quantitatively lower-bound relevant measures of entanglement for such systems.
In particular, we focus on broad classes of variance-based criteria, from which we derive analytical lower bounds to the BSA and the GR. We then use these results to quantify bipartite and multipartite entanglement in nonclassical spin states of atomic ensembles. 

In the experiments we consider, spin-squeezed Bose-Einstein condensates (BECs) of approximately 500 atoms were prepared, and measurements of the collective spin, or of local spin observables were performed after spatially distributing the atoms.
Despite the limited amount of information on the state accessible by such coarse-grained measurements, we show that nontrivial lower bounds on the BSA and the GR can be provided in this case.

Our results constitute a practical method to lower-bound entanglement measures for a variety of physical systems. When applied to atomic ensembles in nonclassical states, this allows one to quantify their usefulness for quantum information tasks beyond metrology, such as quantum teleportation and remote state preparation~\cite{BaoTelep12,KrauterTelep13,ManishRSP20}.

\vspace{2mm}
{\it Preliminaries.---} 
Classes of entanglement measures can be defined as the optimization problem~\cite{brandao05} 
\be\label{eq:witnessedEntDef} 
\EM(\rho) := \max \{ 0 , -\min_{ W \in \mathcal M} \tr[ W \rho] \}  \;,
\ee
where $\mathcal M$ is a subset of entanglement witnesses. The specific choice of this subset results in measures with different interpretations. To give concrete examples, if $\mathcal{W}$ is the set of all entanglement witnesses, choosing $\MBSA=\{W\in\mathcal{W} | \id +W \geq 0 \}$ yields the BSA $\mathcal{E}_M(\rho)=\BSA(\rho)$, while 
choosing $\MGR=\{W\in\mathcal{W} | \id -W \geq 0 \}$ results in the generalized robustness of entanglement $\mathcal{E}_M(\rho)=\GR(\rho)$~\cite{brandao05}.

The idea behind Eq.~\eqref{eq:witnessedEntDef} can be generalized even further, to allow for the definition of entanglement measures that are monotones (\ie cannot increase) only under a subset of LOCC operations. Indeed, we have that:

\vspace{2mm}
{\bf Lemma 1.} {\it Given an operator $K$, the set $\mathcal{M}_{\pm} = \{W\in\mathcal{W} | K\pm W \geq 0 \}$ defines via Eq.~\eqref{eq:witnessedEntDef} an entanglement monotone under LOCC operations commuting with $K$.}

\vspace{2mm}
{\it Proof:} We follow similar arguments as in Refs.~\cite{brandao05,CramerEtAl2013}. Consider some LOCC operation in terms of its Kraus operators $\{ A_k \}_{k}$, with $\sum_k A_k^\dagger A_k \leq \id$. This transforms the state $\rho$ into $\sum_k p_k \rho_k^\prime$, with $p_k:=\tr[A_k \rho A_k^\dagger]$ and $\rho_k^\prime:= A_k \rho A^\dagger_k /p_k$. We now compute
\begin{align}
    &\sum_k p_k\, \EM(\rho_k^\prime) = \sum_k p_k \max \{ 0 , -\tr[W \rho_k^\prime] \} \notag\\
    & \ \ \ = - \sum_i \tr[W A_i\, \rho A_i^\dagger] 
    \,=\, - \sum_i \tr[A_i^\dagger W A_i\, \rho] \notag\\
    & \ \ \ = - \tr[W^\prime \rho] 
    \,\leq\, \EM(\rho), \label{eq:locc}
\end{align}
where the index $i$ runs only over terms $\tr[W A_i \rho A_i^\dagger]<0$, $W^\prime:=\sum_i A_i^\dagger W A_i$, and we used the fact that, if $[A_k,K]=0$, then $0 \leq \sum_i A_i^\dagger (K \pm W) A_i \leq K \pm \sum_i A_i^\dagger W A_i = K \pm W^\prime$, which implies $W^\prime\in\mathcal{M}_{\pm}$.

Equation~\eqref{eq:locc} implies that $\EM$ does not increase on average under the action of LOCC operations commuting with $K$, meaning that it is an entanglement monotone for this subset of LOCC. $\blacksquare$

\vspace{2mm}
For example, for $K=\id$, the resulting measure is a monotone under the full set of LOCC operations, while for $K=\hat{N}$ (the particle number operator) one obtains a monotone under LOCC operations that respect certain superselection rules~\cite{CramerEtAl2013,morris20}. In a concrete situation, however, even if the number of particles fluctuates there always exists an upper bound $\langle\hat{N}\rangle<N$.  
Therefore, using the fact that $\hat{N}\leq N\id$, monotones under the full set of LOCC can always be derived, albeit these might result in lower lower-bounds.

At this point, let us note that: (i) any bounded witness can be rescaled such that $\id \pm W \geq 0$; (ii) the definition in Eq.~\eqref{eq:witnessedEntDef} implies that any witness belonging to $\mathcal M$ delivers a lower bound on $\EM(\rho)$. Therefore, many known witnesses will in general be useful to provide non-trivial lower bounds on entanglement measures. 

In the following, we investigate broad families of entanglement criteria that are experimentally practical and useful, and derive the associated witness operators $W$. This allows us to analytically compute the lower bounds they provide on $\BSA$ and $\GR$, and to apply our approach to experiments with atomic ensembles.

\vspace{2mm}

{\it Bounding entanglement measures from variance-based entanglement criteria.---} 
Let us focus here on classes of entanglement criteria involving variances of operators, and thus involving only measurements of their first and second moments. Because of their simplicity, criteria of this form have been widely investigated in the literature for both bipartite~\cite{hofman03,giovannettietalPRA03,Guhne2004Characterizing,guhnecova,gittsovich08} and multipartite~\cite{SoerensenNAT2001,SoerensenPRL2001,tothPRL07,tothPRA09,vitagliano11,vitagliano14,vitagliano16,MartyVitagliano} scenarios, and they are routinely used experimentally~\cite{PezzeRMP2016}. 

First, let us consider inequalities that are expressed in terms of linear combinations of variances, namely
\be\label{eq:sumCrit}
\mathcal{S}(\rho):= \sum_{k} \var{O_k} - \avg{B} \geq 0
\ee
with $\var{O_k}=\avg{O_k^2}-\avg{O_k}^2$, that hold for all separable states for some operators\footnote{Additional coefficients in front of the variances can be absorbed by redefining $O_k$.} $O_k$ and $B$.

For the sake of simplifying the following discussion, we focus on bounded operators, such that $n^\ast \leq \mathcal{S}(\rho) \leq m^\ast$ for all quantum states. In general, we have
\begin{align}
    n^\ast \geq n &:= \lmax{B}, \label{eq:ndef}\\
    m^\ast \leq m &:= \sum_k \lmax{O_k}^2 - \lmin{B} \label{eq:mdef} \;,
\end{align}
where $\lambda_{\text{min(max)}}(A)$ denoting the minimal (maximal) eigenvalue of $A$. We show here that:

\vspace{2mm}
{\bf Lemma 2.} {\it Every entanglement criterion that can be written in the form of Eq.~\eqref{eq:sumCrit} provides a lower bound on the best separable approximation $\BSA\geq - \mathcal S(\rho)/n$, and to the generalized robustness $\GR\geq - \mathcal S(\rho)/m$.}

\vspace{2mm}
{\it Proof:} For a given state $\rho$, the variance of $O_k$ can be expressed as $\var{O_k} = \min_{s_k} \avg{(O_k - s_k \id)^2}$, where $s_k$ is a real number included in the spectrum of $O_k$ and the minimum is attained for $s_k=\avg{O_k}$.
From this observation, it follows that any criterion in the form of Eq.~\eqref{eq:sumCrit}
can be interpreted as $\mathcal{S}(\rho) = \min_{\bf s} \avg{W(\mathbf{s})}$, which is a minimisation over $\mathbf{s} = \{s_1,, s_2 \dots\}$ of the family of entanglement witness operators
\be\label{eq:Wsproof1}
W(\mathbf{s}) := \sum_{k} ( O_k - s_k\id)^2 - B \;.
\ee
From this definition, it is clear that $W({\bf s})/n \in \MBSA$ and therefore, using Eq.~\eqref{eq:witnessedEntDef}, that $\BSA\geq -\min_{\bf s} \avg{W(\mathbf{s})/n} = - \mathcal {S}(\rho)/n$. 
Similarly, to bound the generalized robustness, one notices that the inequality $\var{O_k} \leq \lmax{O_k}^2$ holds. Therefore, $W({\bf s})/m \in \MGR$, which implies $\GR\geq - \mathcal{S}(\rho)/m$. $\blacksquare$

\vspace{2mm}
As a second relevant class of criteria, we consider inequalities written in the form of modified uncertainty relations, and thus based on the product of two variances. These can be written as
\begin{equation}\label{eq:prodCritU}
\mathcal{U}^2(\rho) := \dfrac{\var{O_1}  \var{O_2}}{\avg{ B}^2} \geq 1
\end{equation}
for all separable states. We now show that:

\vspace{3mm}
{\bf Lemma 3.}{\it
Every entanglement criterion that can be written in the form of Eq.~\eqref{eq:prodCritU} provides a lower bound on the best separable approximation $\BSA\geq \frac{\avg{B}}{n}[1-\mathcal{U}(\rho)]$, and to the generalized robustness.}

\vspace{2mm}
{\it Proof:} First, note that Eq.~\eqref{eq:prodCritU} implies that for all separable states
\begin{equation}\label{eq:prodCrit}
\mathcal{P}(\rho) := \var{O_1}  \var{O_2} - \avg{B}^2 \geq 0  \;.
\end{equation}
This nonlinear inequality can be seen as the result of an optimization over a family of linear inequalities \cite{giovannettietalPRA03}
\begin{align} \label{eq:tangHyp}
\var{O_1}  &\geq 4 \sup_{t\in \mathbb{R}} \left[ \abs{t} \avg{B} - t^2 \var{O_2} \right]  \\
&= - 4 \inf_{t\in \mathbb{R}} \left[ t^2 \var{O_2} - \abs{t} \avg{B} \right] \;, \notag
\end{align}
where $t$ is a real parameter. Geometrically, Eq.~\eqref{eq:prodCrit} can be understood  
as a hyperbola, while Eq.~\eqref{eq:tangHyp} are all its tangents. Note that this procedure is more general than using \eg the triangle inequality $x^2+y^2\geq 2 xy$. 
To summarise, Eq.~\eqref{eq:tangHyp} implies that for any ${t\in \mathbb R}$, all separable states satisfy the inequality $\mathcal{S}_t (\rho):= \var{O_1} + 4 t^2 \var{O_2} - 4 \abs{t} \avg{B} \geq 0$, which takes the form of Eq.~\eqref{eq:sumCrit} with $O_2 \mapsto 2 t O_2$ and $B \mapsto 4 \abs{t} B$. The associated entanglement witness operator is
\begin{equation}
    W({\bf s},t):= (O_1 - s_1 \id)^2 + 4t^2 (O_2 - s_2 \id)^2 - 4 \abs{t} B \;,
\end{equation}
\begin{figure*}[ht!]
  \centering
\includegraphics[width=\textwidth]{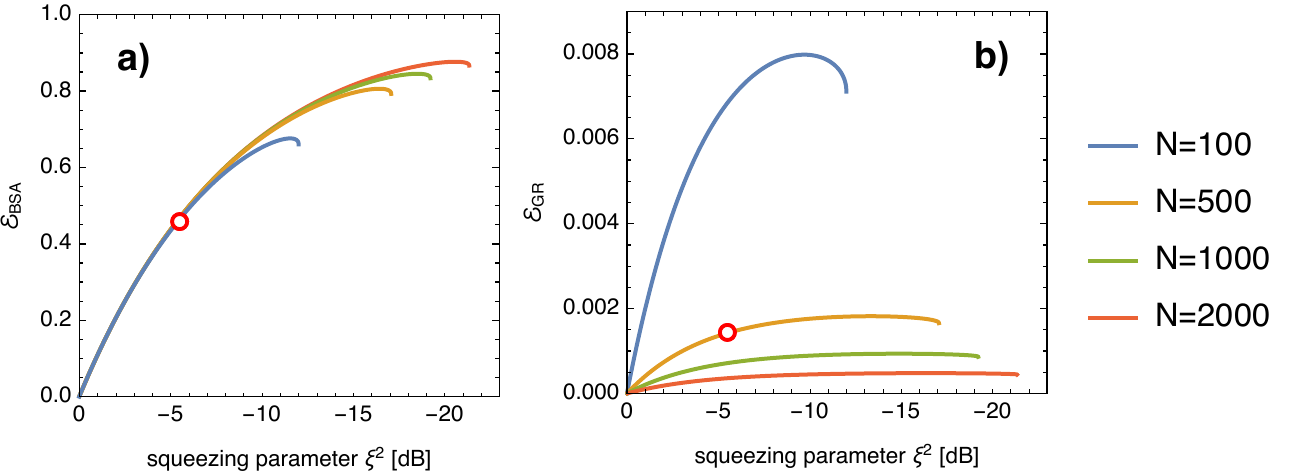}
  \caption{\textbf{Entanglement quantification in spin-squeezed states.} Lower bounds on the best separable approximation, panel \textbf{a}), and on the generalized robustness, panel \textbf{b)}, as obtained from the Wineland spin-squeezing parameter $\xi^2$. Note that for a given number of particles $N$ there is a minimum for $\xi^2$, beyond which the bounds get worse. The red circles correspond to data from measurements of a spin-squeezed BEC with $N=476$ particles.}\label{fig:winBounds}
\end{figure*}
and Eq.~\eqref{eq:prodCrit} can thus be interpreted as $\mathcal{P}(\rho) = \min_{{\bf s},t} \avg{W({\bf s},t)} $. Because $W({\bf s},t)/4\abs{t}n \in \MBSA$, its minimisation gives a lower bound on $\BSA$. This is achieved for $t_\text{BSA}^2=\var{O_1}/4\var{O_2}$, for which we obtain the bound $\BSA\geq -\avg{W({\bf s},t_\text{BSA})/4\abs{t_\text{BSA}}n} = \frac{\avg{B}}{n}[1-\mathcal{U}(\rho)]$.
Similarly, for the generalized robustness we first note that $\mathcal{S}_t \leq m_t := \lmax{O_1}^2+4t^2\lmax{O_2}^2-4t\lmin{B}$. Therefore, $W({\bf s},t)/m_t \in \MGR$ and its minimisation gives a lower bound on $\GR$. This can also be carried out analytically, but since the resulting expressions for $t_\text{GR}^2$ and for $\avg{W({\bf s},t_\text{GR})/m_{t_\text{GR}}}$ are cumbersome, we will not include them here. $\blacksquare$

\vspace{2mm}
In what follows we apply these results to two experimental scenarios that are of broad interest.

\vspace{2mm}
{\it Entanglement quantification in spin-squeezed states.---} As a first application, we quantify multipartite entanglement in a system composed of $N$ spin-$1/2$ particles. An entanglement criterion commonly used in the context of atomic ensembles is based on the Wineland spin-squeezing parameter $\xi^2 := N \var{J_z}/\avg{J_x}^2$~\cite{WinelandPRA1994}, which only requires measurements of the collective spin operator $\bm{J}=\sum_i \bm{\sigma}^{(i)}/2$, where $\bm{\sigma}^{(i)}$ is the vector of Pauli matrices for the $i$th particle. Since $\xi^2\geq 1$ holds for all separable states, observing $\xi^2<1$ certifies entanglement~\cite{SoerensenNAT2001}. This inequality takes the form of Eq.~\eqref{eq:prodCritU} if the constant $N$ in the definition of $\xi^2$ is interpreted as the variance of an operator. 
Following Lemma 3, we obtain the bound
\begin{equation}
    \BSA \geq \mathcal{C} \left( 1 - \sqrt{\xi^2} \right) \;,
\end{equation}
for the BSA, where we have introduced the contrast $\mathcal{C}:=\avg{J_x}/(N/2)$. The exact bound on the GR can also be calculated analytically, and to first order (in $1/N$) it scales as
\begin{equation}
    \GR \geq \dfrac{\mathcal{C}^2}{N}(1-\xi^2) + O(N^{-2}) \;.
\end{equation}
In Fig.~\ref{fig:winBounds} we show the resulting bounds on $\BSA$ and $\GR$ obtained for different particle numbers $N$, and levels of squeezing $\xi^2$.

We now use these tools to quantify entanglement in spin-squeezed BECs of $N=476\pm 21$ ${}^{87}$Rb atoms, magnetically trapped on an atom chip~\cite{SchmiedSCIENCE16}. The two hyperfine states $\left \vert F = 1, m_{F}=-1 \right \rangle \equiv \ket{1}$ and $\left \vert F = 2, m_{F}=1 \right \rangle \equiv \ket{2}$ are identified with a pseudospin $1/2$, such that the entire BEC can be described by a collective spin with $J_z=(N_1-N_2)/2$, \ie half the population difference between the two states. Nontrivial correlations in the system are prepared by controlling atomic collisions with a state-dependent potential~\cite{RiedelNATURE2010}, this give rise to a $J_z^2$ term  in the Hamiltonian 
that results in squeezing of the collective spin state. Atom counting in the two states performed via absorption imaging gives access to a measurement of $J_z$, while measurements along other spin directions are realized by appropriate Rabi rotations before imaging. We obtain $\var{J_z}=32(4)$ and $\mathcal{C}=0.980(2)$, for which $\xi^2=-5.5(6)\,\text{dB}$. The resulting bounds on $\BSA$ and $\GR$ are shown as red circles in Fig.~\ref{fig:winBounds}.

\begin{figure*}[ht!]
  \centering
\includegraphics[width=\textwidth]{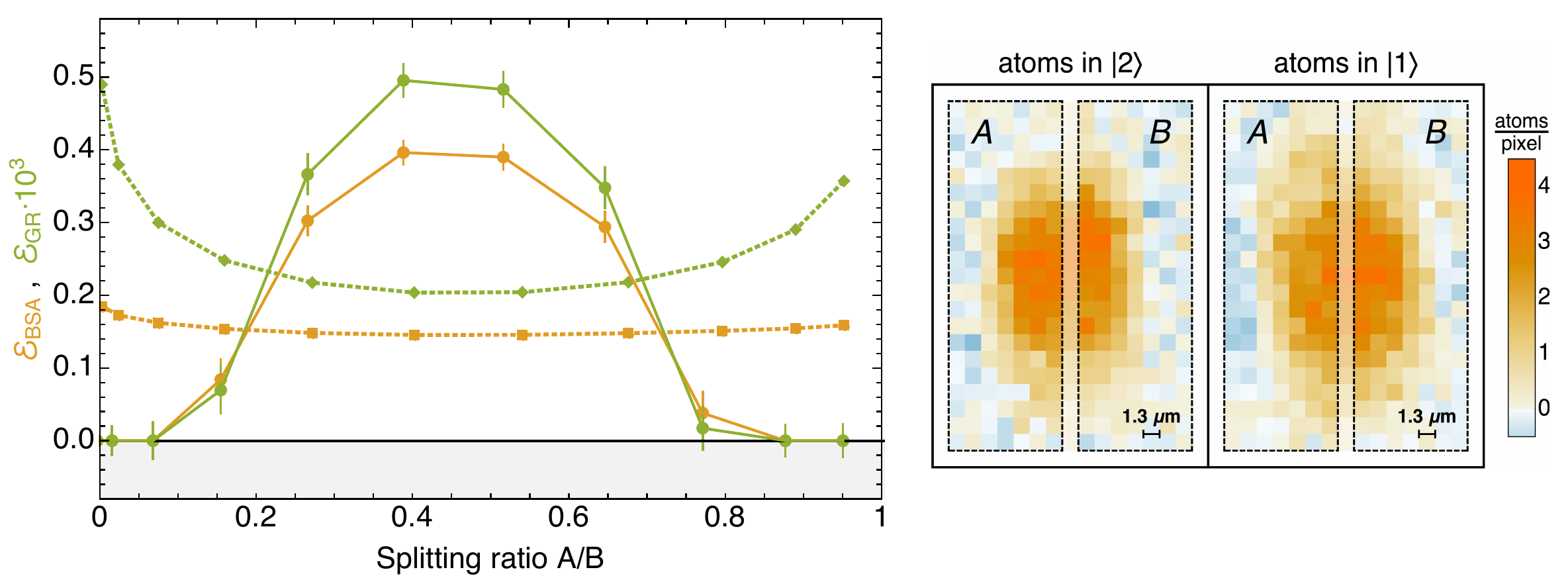}
  \caption{\textbf{Entanglement quantification in split spin-squeezed BECs.} Lower bounds on the BSA and the GR, as obtained from Eq.~\eqref{eq:SGiova} according to Lemma 3. Measurements are taken from a spin-squeezed BEC of $N=590$ atoms. The dotted lines show the maximum amount of entanglement that could be explained by detection cross-talk~\cite{Fadel18}. On the right, we show single-shot absorption images of the atomic densities for the two internal degrees of freedom, with an example of regions $A$ and $B$ used to define the collective spins $J^A$ and $J^B$.
  }\label{fig:gioBounds}
\end{figure*}

\vspace{2mm}
{\it Entanglement quantification in split spin-squeezed states.---} As a second application, we quantify bipartite entanglement between two systems, $A$ and $B$. An entanglement criterion suitable to this scenario is the one derived by Giovannetti \textit{et al.} in Ref.~\cite{giovannettietalPRA03}, stating that all separable states of two collective spins satisfy
\begin{equation}
\mathcal{G}^2:= \dfrac{\var{g_z J_z^A + J_z^B} \var{g_y J_y^A + J_y^B} }{\left(|g_zg_y| \abs{\avg{J_x^A}} + \abs{\avg{J_x^B}} \right)^2 /4}  \geq 1  \label{eq:SGiova}
\end{equation}
for $g_z,g_y\in\mathbb{R}$. The latter variables parametrize a family of inequalities of the form of Eq.~\eqref{eq:prodCritU}, and hence Lemma 3 applies. The larges lower bound on $\BSA$ and $\GR$ arises from a minimization of Eq.~\eqref{eq:SGiova} over $g_z$ and $g_y$.

We now use these tools to quantify entanglement between two partitions of a spin-squeezed BEC. For this, atoms are spatially distributed before performing high-resolution absorption imaging. Then, we define spatially separated regions on the images, and associate them to a local spin that is measured by counting the local population difference in the two hyperfine states~\cite{Fadel18}, see Fig.~\ref{fig:gioBounds}~(right). We consider BECs of $N=590\pm 30$ atoms, showing a squeezing of $\xi^2=-3.8(2)\,\text{dB}$, and investigate different bipartitions by splitting the images horizontally into two parts. In Fig.~\ref{fig:gioBounds}~(left) we plot the lower bounds on $\BSA$ and $\GR$ obtained from Eq.~\eqref{eq:SGiova}.

\textit{Conclusions.---} In this work we presented a practical method to lower-bound classes of entanglement measures from specific families of entanglement witnesses. In particular, we give analytical lower bounds on two relevant measures, the best separable approximation (BSA) and the generalized robustness (GR), as a function of the observed violation of entanglement criteria that are routinely used experimentally. Remarkably, this approach can provide non-trivial bounds even when very limited amount of information on the state is available. 

To illustrate the usefulness our method, we give two concrete applications of entanglement quantification in atomic ensembles. In the first, we consider a spin-squeezed BEC, and show that measurements of the collective spin length and squeezed quadrature are sufficient to lower-bound the BSA and GR. This allows us to relate these measures to the Wineland spin-squeezing coefficient associated to the metrological usefulness of a quantum state. In the second application, we quantify the bipartite entanglement observed between two atomic ensembles. By spatially distributing a spin-squeezed BEC we are able to define two local collective spins, each associated with a different spatial region, whose joint state violates an entanglement criterion. From our results, we are able to translate the value of this violation into lower bounds on the BSA and the GR.

Our investigation opens up new possibilities to quantify and characterize entanglement in atomic ensembles. Apart from their fundamental interest, our results could be useful for concrete applications in quantum technologies, in particular with regards to entanglement-based benchmarking of complex quantum systems. There, certification and quantification of entanglement implies both appropriate levels of coherence (hence the system’s quantum-nature) and the ability to fully control the system, thus allowing one to test the functionality of quantum devices.

\vspace{2mm}
\textit{Acknowledgments.---} We thank G\'eza T\'oth, Pavel Sekatski for discussions and Philipp Treutlein for giving access to experimental data. M.F. was partially supported by the Research Fund of the University of Basel for Excellent Junior Researchers. A.U. appreciates the hospitality and support from IQOQI-Vienna during her visit and acknowledges financial support from OIST Graduate University and from a Research Fellowship of JSPS for Young Scientists. 
M.H., N.F. and G.V. acknowledge support from the Austrian Science Fund (FWF) through projects Y879-N2 (START), P 31339-N27 (Stand-alone), ZK 3 (Zukunftskolleg) and M 2462-N27 (Lise-Meitner).

\newpage
\bibliographystyle{apsrev4-1fixed_with_article_titles_full_names_new}
\bibliography{bibliography}

\end{document}